\documentclass[aps,jcp,twocolumn,superscriptaddress]{revtex4-2}
\usepackage{amsmath}
\usepackage{amssymb}
\usepackage{color}
\usepackage{graphicx}
\usepackage{epstopdf}

\DeclareMathOperator{\artanh}{artanh}

\begin{document}

\title{Surface Charge--Potential Relation for Spherical Particles in Electrolyte Solutions}
\author{Olga I. Vinogradova}
\email[Corresponding author: ]{oivinograd@yahoo.com}
\affiliation{Frumkin Institute of Physical Chemistry and Electrochemistry, Russian Academy of Sciences, 31-4 Leninsky Prospect, 119071 Moscow, Russia}
\author{Elena F. Silkina}
\affiliation{Frumkin Institute of Physical Chemistry and Electrochemistry, Russian Academy of Sciences, 31-4 Leninsky Prospect, 119071 Moscow, Russia}

\begin{abstract}
Predicting the relationship between surface charge density and electrostatic potential for spherical particles, known as the contact theorem, remains a fundamental challenge in colloid science. Because the governing non-linear Poisson--Boltzmann equation lacks an analytical solution, researchers typically rely on numerical calculations or various semi-empirical approximations. In this paper, we overcome these limitations by developing a dual-asymptotic framework that provides explicit, closed-form expressions for this relationship across the entire spectrum of particle curvature. For weakly curved systems ($\lambda_D/R \ll 1$), where the Debye length $\lambda_D$ is much smaller than the particle radius $R$, a formal mathematical derivation rigorously establishes the Ohshima--Healy--White formula as an exact regular perturbation expansion. Conversely, for highly curved spheres ($R/\lambda_D \ll 1$), we employ singular perturbation analysis. This approach provides a justification for the widely used \emph{ad hoc} linearization, proving that the spherical Debye--Hückel contact theorem remains asymptotically exact within the full Poisson--Boltzmann theory at finite surface potentials due to the geometric deactivation of non-linearity. Crucially, we map the exact limits of this geometric regulation, demonstrating how non-linear screening re-emerges as the particle radius increases, with the breakdown threshold governed by the interplay between curvature and surface charge density.
\end{abstract}

\maketitle

\section{Introduction}\label{sec:intro}

When solids are suspended in an electrolyte solution of concentration $n_{\infty}$, they typically assume certain surface charge densities $\sigma$ and surface potentials $\Phi_s$ that are linked to each other. For non-conducting materials, $\sigma$ remains constant and is independent of salt concentration. Importantly, many surface phenomena are driven by $\Phi_s$ and not, as one might initially think, by surface charge density and do not depend on the potential profile near the surface. The best-known examples include the limiting value of the ionic density at the boundary~\cite{andelman.d:2006} and electrokinetic phenomena~\cite{overbeek.jtg:1950}. Consequently, once the relation between $\sigma$ and $\Phi_s$ is known, the limiting osmotic pressure, macroscopic electrokinetic response, or differential capacitance can be found without analyzing the full potential profile in the solution.

A closed-form analytical relationship between $\sigma$ and $\Phi_s$, termed the contact theorem (and also the boundary relationship), can be obtained by the first integration of the second-order Poisson--Boltzmann equation. However, due to its non-linearity, exact solutions exist only for a limited number of simple configurations. For a planar wall, the exact solution is given by the Grahame equation~\cite{grahame.dc:1947,andelman.d:2006,herrero.c:2024}. Some approximate expressions are also available for the confined geometry of nanochannels~\cite{silkina.ef:2019}. Nevertheless, the experimentally relevant geometry of spherical particles of radius~$R$ remains challenging, and not even an explicit first integral exists. Closed-form rigorous solutions are known only for the Debye--H\"{u}ckel regime of low surface potentials and for infinitely dilute solutions, where the Poisson--Boltzmann equation reduces to the Laplace equation; outside these limits, numerical approaches that present their own challenges have mostly been followed.

There is a large literature describing the electrostatic properties of dispersions of highly charged particles in an electrolyte solution (colloidal suspensions) and the idealized case of a single sphere, which represents the limit of infinite particle dilution. Theoretical studies have traditionally focused on spatial potential profiles---which inevitably require the second integration of the Poisson--Boltzmann equation---or far-field properties, while the surface behavior remains less explored. Below, we mention what we consider to be the most relevant contributions.

During the last decades, several theoretical papers have been concerned with the description of the far-field electrostatics using the so-called effective (renormalized) charge. A comprehensive discussion of this effective charge paradigm, which now stretches from the cell model of \citet{alexander.s:1984} and the saturation limits of \citet{ramanathan.g:1986, ramanathan.g:1988} to the multi-scale perturbation schemes of \citet{trizac.e:2002} and the high-order quasi-planar expansions of~\citet{samaj.l:2015}, can be found in several review articles~\cite{belloni.l:1998,levin.y:2002,andelman.d:2006}. While these global approaches appear to be successful in describing the collective behavior of suspensions, they inherently adopt a \textit{top-to-bottom} perspective by analyzing the linear Debye--H\"{u}ckel tails far from the spheres and do not connect the results with the exact surface electrostatics. Thus, the contact theorem for curved geometries formulated by \citet{mallarino.jp:2015} within the cell model concerns the long-distance electrostatic coupling, and the associated effective surface potential does not coincide with $\Phi_s$, as emphasized already by \citet{belloni.l:1998}. Consequently, deriving an explicit relation between $\sigma$ and the true surface potential $\Phi_s$---the curved-surface generalization of the Grahame equation---is a distinct problem that cannot be resolved within the effective charge concept.

Some related \textit{bottom-to-top} analytical constructions attempted to derive approximate expressions for the potential over the entire diffuse layer. \citet{shkel.i:2000} obtained a quasi-planar approximation in the form of functional series, which are difficult to use and have limited predictive capabilities. Later, \citet{tuinier.r:2003} proposed a practical analytical approximation for the potential profile around spherical particles by adapting the exact planar solution. While convenient for calculations, this expression relies on an interpolation ansatz rather than a rigorous derivation. These authors, however, made no attempt to connect the results with the surface charge--potential relationship, and despite broader claims, their approaches remain fundamentally restricted to the limit of small curvature.

For large, weakly curved particles ($R/\lambda_D\gg1$, where $\lambda_D$ is the Debye length of the electrolyte solution), the foundational numerical benchmark was established in the tabulated work of~\citet{loeb.a:1961}. Early attempts to construct a rigorous perturbation approach capable of reproducing these numerical results were pioneered by~\citet{white.lr:1977}, although the resulting analytical solution proved too cumbersome for practical use. To circumvent this difficulty, \citet{ohshima.h:1982} subsequently derived a more tractable boundary relation by evaluating the leading curvature correction approximately from the planar profile, arguing that it holds to first order in $\lambda_D/R$ for arbitrary surface potentials. From the perspective of formal perturbation theory, however, such a construction bridges the non-linear planar solution and spherical geometry without a systematic derivation, effectively acting as an interpolation ansatz.
Later, some asymptotic methods were developed for this limit. For instance, \citet{natarajan.r:1984} applied matched asymptotic expansions to the non-linear Poisson--Boltzmann equation around rigid spheres. While such singular perturbation schemes are required to construct a uniformly valid potential profile across the entire infinite domain, they are unnecessary in the weakly curved regime for deriving the contact theorem, which depends solely on the local potential gradient at the surface. Moreover, because their analysis relied on cumbersome higher-order series, it obscured a simple, elementary derivation of the first-order Ohshima--Healy--White curvature correction. Consequently, a transparent derivation of this relation via a direct regular perturbation expansion has remained elusive.

This missing foundation becomes even more consequential in the opposite limit of highly curved, ultra-small particles ($R/\lambda_D \ll 1$), which remains largely unexplored theoretically, although it dates back to the famous theory of electrolytes by~\citet{debye.p:1923}. Being practically minded, these authors simply linearized the Poisson--Boltzmann equation, although this is only justified at small potentials. According to~\citet{onsager.l:1933}, retaining the full Boltzmann factor within the non-linear framework violates statistical self-consistency and leads to a strong short-range divergence for point charges, so linearization is a necessary step. However, small particles (and even ions~\cite{kadhim.mj:2020}) in solution have a finite radius and carry an effective surface charge density, so no point-charge singularity arises and the full non-linear equation can be solved directly. Yet numerical calculations of the potential profiles around tiny inorganic ions~\cite{vinogradova.oi:2023,vinogradova.oi:2025} revealed that these exact non-linear solutions closely recover the linearized Debye--H\"{u}ckel predictions even though the linear theory is conventionally expected to fail at such high potentials. Currently, the theoretical origin of this asymptotic convergence, already anticipated in the old colloid literature, is still poorly understood.
 This difficulty in obtaining analytical insight stems from the inherent mathematical complexity of the problem. Here, standard regular expansions fail because the intense local field near the interface is coupled to the global electrolyte screening cloud. The problem is intrinsically multiscale, involving an inner region governed by the particle radius $R$ and an outer region scaled by the Debye length $\lambda_D$, which requires singular perturbation approaches. While~\citet{macgillivray.ad:1969} applied matched asymptotic expansions in the limit of infinite dilution ($R/\lambda_D \to 0$), his analysis was restricted to the leading order and relied on an ad hoc choice of the matching constant to recover the Debye--H\"{u}ckel expression. Later, \citet{natarajan.r:1986} pursued higher-order expansions for constant-potential spheres using cumbersome series and ad hoc semi-heuristic modifications, which broke down already at moderate potentials. Although their leading terms recovered the Debye--H\"{u}ckel law, no attempt was made to establish its explicit critical threshold or to address constant-charge particles, where the surface potential adjusts self-consistently. Instead, this was claimed to be true for any value of the surface potential (and charge).

In this paper, we develop a dual-asymptotic framework that systematically resolves the boundary electrostatics at both curvature extremes. By constructing a rigorous regular expansion, we prove that the Ohshima--Healy--White expression is the exact asymptotic result of the non-linear spherical Poisson--Boltzmann equation for weak curvatures. Even more significantly, we then provide a straightforward derivation of a matched, closed-form boundary relation for highly curved, constant-charge particles, demonstrating that the spherical Debye--H\"{u}ckel contact relation naturally emerges as the mathematically exact first-order asymptotic limit of the full non-linear problem at small $R/\lambda_D$. The analysis reveals that, although the far-field electrostatics is effectively blinded by the non-linear charge saturation captured by global effective-charge theories, the local surface potential remains governed by the particle's true surface charge density. This regulated response suppresses the otherwise expected exponential non-linearity up to a well-defined critical threshold, for which we establish an explicit validity criterion. Within this regime, such a geometric deactivation of mean-field non-linearity at high curvature reproduces the classical Debye--H\"{u}ckel boundary relationship even at high surface potentials.

Methodologically, our work establishes a rigorous \textit{bottom-to-top} (near-field to far-field) framework that stands in sharp contrast to the \textit{top-to-bottom} approaches historically favored in the statistical physics of charged colloids. Traditional theoretical treatments often start their analysis from the far-field asymptotic limit that guarantees a weak potential, and then move downward toward the particle. When faced with the short-range escalation of the Boltzmann factor, these \textit{top-to-bottom} models are frequently forced to introduce effective concepts such as charge renormalization to keep the equations tractable. By contrast, our regular and singular perturbation approach that proceeds directly from the sphere's surface proves that the intrinsic three-dimensional geometry of the sphere naturally tames the non-linearities from the \textit{bottom-to-top}, smoothly matching the surface conditions with the bulk electrolyte solution.

The remainder of the paper is arranged as follows. In Sec.~\ref{sec:framework}, we introduce the model and derive the exact integral relation between the surface charge and surface potential. Section~\ref{sec:analytics} describes the classical solutions that define the leading-order limits of our subsequent perturbation expansions. In Sec.~\ref{sec:expansions}, we derive explicit charge--potential relations for the low- and high-curvature cases. Section~\ref{sec:results} compares these predictions with the numerical solutions of the full non-linear equation. We conclude in Sec.~\ref{sec:concl}. Appendix~\ref{app:equivalence} examines the crossover from the high-curvature linear regime to the non-linear screening regime.

\section{Theoretical Framework and Non-Linear Charge-Potential Problem}\label{sec:framework}

\begin{figure}[h]
	\includegraphics[width=1.0\columnwidth]{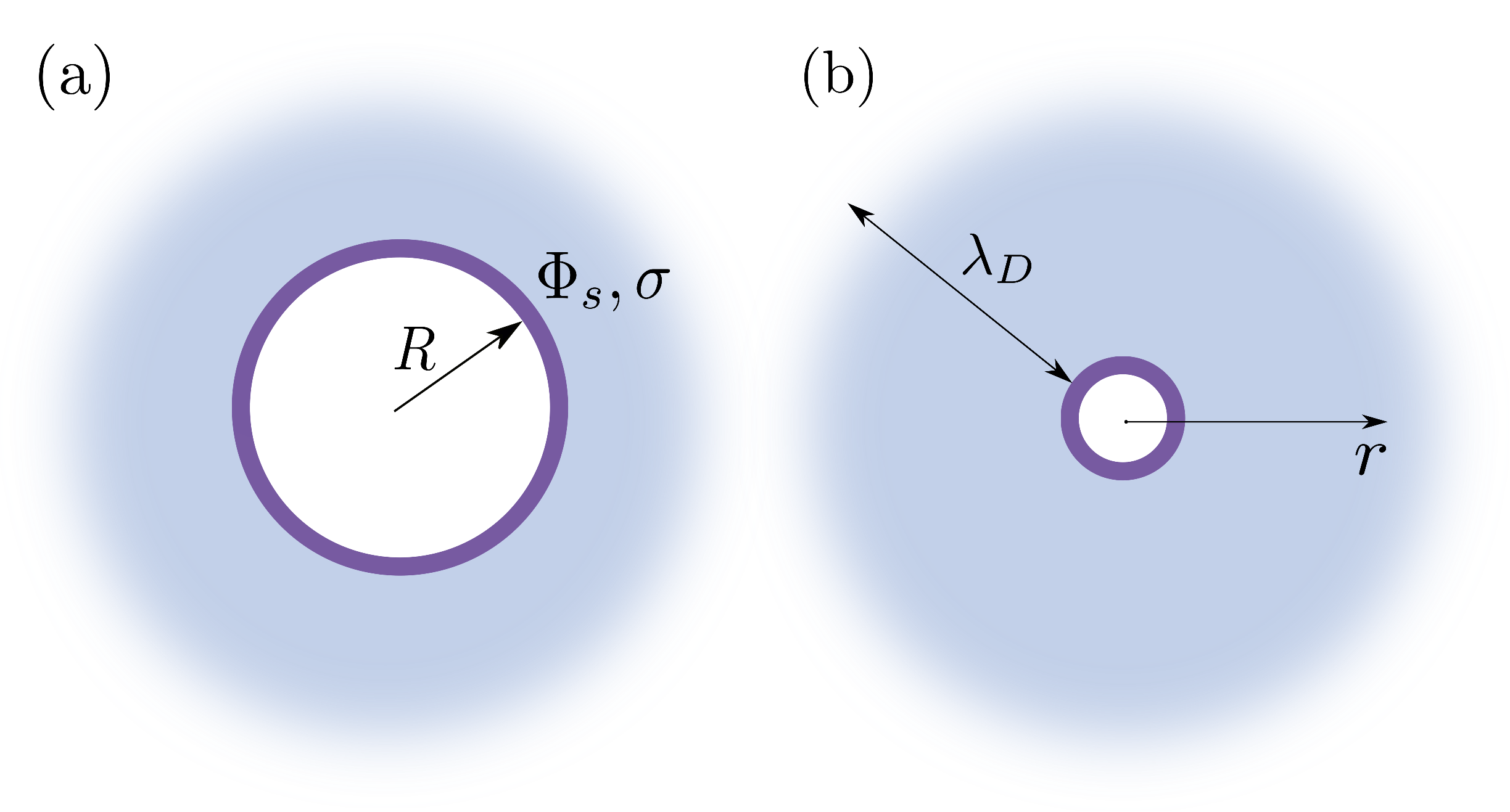}
	\caption{Schematic representation of charged large (a) and small (b) spherical particles of radius $R$ in electrolyte solutions. The surrounding ionic clouds of thickness $\lambda_D$ are shown by colored areas. The surface of the spheres assumes a charge density $\sigma$ and a potential $\Phi_s$. }
	\label{fig:sketch}
\end{figure}

We consider a  charged spherical particle of radius $R$ immersed in a bulk symmetric 1:1 electrolyte solution with a static dielectric permittivity $\varepsilon$. The sphere surface  ($r=R$) is  characterized by a potential $\Phi_s$ and a charge density $\sigma$. In response to the surface charge, an electrostatic diffuse layer (EDL) forms around the particle. Its thickness is governed by the bulk Debye screening length $\lambda_D$, as schematically illustrated in Fig.~\ref{fig:sketch}.

The electrostatic potential $\Phi(r)$ satisfies the non-linear spherical Poisson–Boltzmann equation:
\begin{equation}\label{eq:SPB}
\frac{1}{r^2}\frac{d}{dr}\left(r^2 \frac{d\phi}{dr}\right) = \lambda_D^{-2} \sinh\phi,
\end{equation}
where $\phi = e|\Phi|/(k_B T)$ is the dimensionless potential, $e$ is the elementary charge, $k_{B}$ is the Boltzmann constant, and $T$ is the absolute temperature. The Debye screening length is defined as $\lambda_{D}=\left( 8\pi \ell _{B}n_{\infty}\right) ^{-1/2}$, where $n_\infty$ is the bulk ion number density, and $\ell _{B}=e^{2}/(\varepsilon k_{B}T)$ denotes the Bjerrum length of the solvent with $\varepsilon$ representing the static permittivity. To satisfy global electroneutrality, this diffuse layer contains a total compensating counterion charge of equal magnitude and opposite sign to that of the particle.

To solve Eq.~\eqref{eq:SPB}, we impose the far-field boundary conditions:
\begin{equation}\label{eq:infinity}
\phi(\infty) = 0, \quad \left.\frac{d\phi}{dr}\right|_{r=\infty} = 0,
\end{equation}
along with the surface conditions at $r=R$, specifying  a constant surface charge density (Neumann):
\begin{equation}\label{eq:CC}
\left.\frac{d\phi}{dr}\right|_{r=R} = -\frac{2}{\ell_{GC}},
\end{equation}
where $\ell_{GC} = e/(2\pi \ell_B |\sigma|)$ denotes the Gouy–Chapman length.
The  surface potential,
\begin{equation}\label{eq:CP}
\phi_s = \phi(R),
\end{equation}
is then set up self-consistently. Our purpose here is to derive analytical equations that relate $\ell_{GC}$ with $\phi_s$ in the limit of low and high surface curvature. Such an equation is referred below to as boundary relation or the contact theorem.

The contact relation is obtained by integrating Eq.~\eqref{eq:SPB} once, which is beset with difficulties. Multiplying both sides by $r^2$ and integrating from the particle surface $r=R$ to infinity ($r\to\infty$) yields, upon applying the boundary conditions~\eqref{eq:infinity}:
\begin{equation}\label{eq:first_int_exact}
-R^2 \left.\frac{d\phi}{dr}\right|_{r=R} = \lambda_D^{-2} \int_R^\infty r^2 \sinh\phi(r) \,dr.
\end{equation}
Substituting the Neumann condition~\eqref{eq:CC} into the left-hand side of Eq.~\eqref{eq:first_int_exact} leads to the master integral equation:
\begin{equation}\label{eq:integral_master}
\frac{\lambda_D}{\ell_{GC}} = \frac{1}{2R^2\lambda_D} \int_R^\infty r^2 \sinh\phi(r) \,dr.
\end{equation}
Physically, Eq.~\eqref{eq:integral_master} expresses global electroneutrality. The integral on the right-hand side represents the total screening charge accumulated within the electrostatic diffuse layer, which balances the net surface charge of the particle on the left-hand side.
Although Eq.~\eqref{eq:integral_master} is exact, it cannot be solved in closed form because the integrand explicitly depends on the unknown potential profile $\phi(r)$. This integral relation highlights a defining mathematical feature of the spherical problem: the particle radius $R$ enters explicitly as both a geometric prefactor and the lower integration limit. Consequently, any variation in particle size fundamentally reshapes the link between the surface charge and potential.

In the limit $R \to \infty$, Eq.~\eqref{eq:integral_master} must smoothly recover the classic planar Grahame relation:
\begin{equation} \label{eq:grahame}
\frac{\lambda_D}{\ell_{GC}} = \sinh\left(\frac{\phi_s}{2}\right).\tag{G}
\end{equation}

The behavior in the highly curved limit ($R/\lambda_D \to 0$) requires special comment. Shrinking the particle radius toward the point-charge limit might naively be expected to trigger the catastrophic exponential divergences first diagnosed by~\citet{onsager.l:1933} within the full Poisson--Boltzmann theory. However, a closer inspection of Eq.~\eqref{eq:integral_master} reveals that for a finite-radius sphere, the exact non-linear equation automatically regulates this geometric coupling through a curvature-induced potential suppression. This can be rationalized as follows.

Because the potential profile $\phi(r)$ decays monotonically away from the surface, it is bounded from above by its surface value, $\phi(r) \le \phi_s$. Furthermore, since $\sinh\phi > \phi$ holds for any non-vanishing potential, the true non-linear screening charge in Eq.~\eqref{eq:integral_master} is strictly bounded from below by its linearized counterpart:
\begin{equation}\label{eq:app_ineq}
\frac{\lambda_D}{\ell_{GC}} > \frac{1}{2R^2\lambda_D} \int_R^\infty r^2 \phi(r) \,dr.
\end{equation}
Under a fixed surface charge density ($\ell_{GC} = \text{const}$), the left-hand side remains invariant. In the high-curvature regime ($R/\lambda_D \ll 1$), integrating the right-hand side by parts isolates the leading-order geometric contribution. For the inequality to hold despite the divergence of the prefactor $1/(2R^2\lambda_D)$, the surface potential must be strongly suppressed and scale as $\phi_s \propto R/\lambda_D$. This geometric constraint inherently limits the potential near the surface, preventing non-linear exponential growth and forcing the full non-linear Poisson--Boltzmann equation into an effectively linear regime, even for highly charged particles. Beyond a certain electrostatic threshold, however, this linear scaling breaks down as strong charge accumulation contracts the screening cloud, forcing the potential profile back toward the non-linear Grahame limit.

\section{Analytical Limits}\label{sec:analytics}

Before turning to the full dual-asymptotic framework, it is instructive to outline the classical limits where the governing equation simplifies immediately. These special cases provide the exact reference points that define the leading-order limits of our subsequent perturbation expansions.

\subsection{The Laplacian Limit}	\label{subsec:laplacian}

When $R/\lambda_D$ is extremely small, which is the case of nanoparticles and ions in highly dilute solutions, the r.h.s. of Eq.~\eqref{eq:SPB} can be neglected. In this case Eq.~\eqref{eq:SPB} reduces to the Laplace equation:
\begin{equation}  \label{eq:L_sphere}
\frac{1}{r^2} \frac{d}{d r} \left( r^2 \frac{d \phi}{d r} \right) \simeq 0.
\end{equation}
Solving Eq.~\eqref{eq:L_sphere} by enforcing the far-field conditions \eqref{eq:infinity} and either the Dirichlet~\eqref{eq:CP} or the Neumann~\eqref{eq:CC} constraint  at the surface yields the following potential profile:
\[
\phi(r) \simeq \phi_s \frac{R}{r} \simeq \frac{2 R}{\ell_{GC}} \frac{R}{r}.
\]
This is the unscreened Coulomb potential of a point charge. It follows then that the boundary relation is given by:
\begin{equation}\label{eq:phil_sp}
\boxed{\phi_s = \frac{2 R}{\ell_{GC}}} \tag{L}.
\end{equation}
Thus, the fixed surface charge density determines the surface potential, which becomes constant, i.e., independent of salt.

\subsection{The Linearized Limit (Debye–Hückel Theory)}

If $\phi_s \le 1$, $\sinh\phi \simeq \phi$ and Eq.~\eqref{eq:SPB} can be linearized to give:
\begin{equation}  \label{eq:DH_sphere}
\frac{1}{r^2} \frac{d}{d r} \left( r^2 \frac{d \phi}{d r} \right) \simeq \lambda_D^{-2} \phi.
\end{equation}
The solution to Eq.~\eqref{eq:DH_sphere} satisfying \eqref{eq:CP} and \eqref{eq:infinity} is well known and reads:
\[
\phi(r) \simeq \phi_s \frac{R}{r} e^{-(r-R)/\lambda_D}.
\]
Substituting this profile into the exact integral relation \eqref{eq:integral_master} and making standard manipulations, we obtain:
\begin{equation}\label{eq:contact_DH}
\boxed{\frac{\lambda_D}{\ell_{GC}} \simeq \frac{\phi_s}{2}\left(1 + \frac{\lambda_D}{R}\right)}. \tag{DH}
\end{equation}

We emphasize that Eq.~\eqref{eq:contact_DH} is derived assuming the surface potential is low, but no restrictions concerning the curvature are imposed. For a plane ($R \to \infty$), we recover $\phi_s \simeq 2\lambda_D/\ell_{GC}$, which represents the linearized planar relation. We thus conclude that the curvature adds a correction proportional to $\lambda_D/R$. If $\lambda_D/R$ is extremely large, Eq.~\eqref{eq:contact_DH} reduces to \eqref{eq:phil_sp}.

\section{Contact theorem via regular and singular expansions}\label{sec:expansions}

In this section, we derive the relation between $\phi_s$ and $\lambda_D/\ell_{GC}$ by using the perturbation approaches.

\subsection{Low Curvature Expansion (Large Sphere)}\label{subsec:low_curvature}

When the sphere is large, meaning its radius is much larger than the EDL thickness, the natural parameter is $\epsilon^{-1} = \lambda_D/R \ll 1$, where $\epsilon = R/\lambda_D$. The solution can then be expanded in powers of $\epsilon^{-1}$:
\begin{equation}\label{eq:series_phi}
  \phi = \phi_0 + \epsilon^{-1}\phi_1 +  \mathcal{O}(\epsilon^{-2}),
\end{equation}

Let us introduce the dimensionless variables
\[
\xi = \frac{r-R}{\lambda_D}, \quad \rho = \frac{r}{R},
\]
where $\xi$ is a stretched radial coordinate and $\rho$ is kept as a slow variable. Then $\phi = \phi(\xi,\rho)$ and
\begin{equation}\label{eq:r}
  r = R (1 + \epsilon^{-1} \xi).
\end{equation}
Since $\xi$ and $\rho$ are treated as independent to leading order within the multi-scale framework, the total radial derivative operator can be written as
\[
\frac{d}{dr} = \frac{1}{\lambda_D}\frac{\partial}{\partial\xi} + \frac{1}{R}\frac{\partial}{\partial\rho}.
\]
In terms of $\epsilon^{-1}$, this operator reads:
\[
\frac{d}{dr} = \frac{1}{R}\left(\epsilon\frac{\partial}{\partial\xi} + \frac{\partial}{\partial\rho}\right).
\]
Substituting this operator into Eq.~\eqref{eq:SPB} and using Eq.~\eqref{eq:r}, we find after some algebra that the potential profile obeys:
\begin{equation}\label{eq:PB_low_curvature_operator}
\begin{split}
  \epsilon^2\frac{\partial^2\phi}{\partial\xi^2}
  &+ \epsilon\left( \frac{2}{1+\epsilon^{-1}\xi}\frac{\partial\phi}{\partial\xi}
  + 2\frac{\partial^2\phi}{\partial\xi\partial\rho} \right) \\
  &+ \mathcal{O}(1) = \epsilon^2 \sinh\phi,
\end{split}
\end{equation}
where the expansion for $\phi$ is given by Eq.~\eqref{eq:series_phi}. Equating powers of $\epsilon^2$ and $\epsilon$ yields the hierarchy of differential equations discussed below.

Collecting terms of order $\epsilon^2$ yields
\[
	\frac{\partial^2\phi_0}{\partial\xi^2} = \sinh\phi_0.
\]
This is the planar Poisson--Boltzmann equation. Its solution satisfying $\phi_0(0,\rho) = \phi_s$ and $\phi_0(\infty,\rho)=0$ is independent of $\rho$. Performing the first integration we obtain
\begin{equation}\label{eq:LC_planar}
		\frac{1}{2}\left(\frac{\partial\phi_0}{\partial\xi} \right)^2 = \cosh\phi_0 - 1, \quad \left.\frac{\partial\phi_0}{\partial\xi}\right|_{\xi=0} = -2 \sinh\frac{\phi_s}{2}.
\end{equation}

Hence
\[
	\phi_0(\xi) = 4\artanh\!\left(e^{-\xi}\tanh\frac{\phi_s}{4}\right).
\]

Collecting terms of order $\epsilon$, we derive:
\[
\frac{\partial^2\phi_1}{\partial\xi^2} - \cosh\phi_0\,\phi_1
= -2\frac{\partial^2\phi_0}{\partial\xi\partial\rho} - \frac{2}{1+\epsilon^{-1}\xi}\frac{\partial\phi_0}{\partial\xi}.
\]
Since $\partial\phi_0/\partial\rho = 0$, the first term on the r.h.s. vanishes. Expanding $(1+\epsilon^{-1}\xi)^{-1} = 1 - \epsilon^{-1}\xi + \cdots$ and collecting terms strictly of order $\epsilon$ transforms this into the linear differential equation:
\begin{equation}\label{eq:second_term}
\frac{\partial^2\phi_1}{\partial\xi^2} - \cosh\phi_0\,\phi_1 = -2\frac{\partial\phi_0}{\partial\xi},
\end{equation}
subject to the boundary conditions $\phi_1(0,\rho) = 0$ and $\phi_1(\infty,\rho)=0$.

To find $\partial \phi_1/\partial \xi$, we multiply Eq.~\eqref{eq:second_term} by $\partial\phi_0/\partial\xi$ and integrate from $0$ to $\infty$. Using integration by parts and taking into account the boundary conditions, we obtain:
\[
\left.\frac{\partial\phi_1}{\partial\xi}\right|_{\xi=0} = \frac{2}{\left(\frac{\partial\phi_0}{\partial\xi}\right)_{\xi=0}} \int_0^\infty \left(\frac{\partial\phi_0}{\partial\xi}\right)^2 d \xi.
\]
Substituting Eqs.~\eqref{eq:LC_planar}, this expression simplifies to:
\[
\left.\frac{\partial\phi_1}{\partial\xi}\right|_{\xi=0} = - \frac{2\int_0^\infty (\cosh\phi_0 - 1) d \xi}{\sinh\frac{\phi_s}{2}}.
\]
Using the relation $d\xi = -d\phi_0 / \sqrt{2(\cosh\phi_0 - 1)}$ to change the integration variable from $\xi$ to $\phi_0$, this spatial integral evaluates analytically, yielding the first-order boundary derivative exactly:
\begin{equation}\label{eq:phi1_prime_final}
\left.\frac{\partial\phi_1}{\partial\xi}\right|_{\xi=0} = - \frac{4\left(\cosh\frac{\phi_s}{2}-1\right)}{\sinh\frac{\phi_s}{2}} = -4\tanh\frac{\phi_s}{4}.
\end{equation}

While this non-linear curvature dependence was originally introduced by~\citet{ohshima.h:1982}, our approach derives the $-4\tanh(\phi_s/4)$ term through a rigorous, self-consistent regular perturbation expansion of the non-linear Poisson--Boltzmann equation. Notably, this direct regular framework yields closed-form analytical results without partitioning the domain into intermediate matching zones~\cite{erker.j:1996}. Physically, this first-order boundary correction (proportional to $\epsilon^{-1}$) saturates at high potentials.
This saturation is physically consistent: for highly charged surfaces, the intense counterion accumulation clamps the screening cloud extremely close to the interface. Because the effective thickness of this inner diffuse layer becomes much smaller than the particle radius, the local electrostatic field becomes insensitive to the geometric curvature, and the relative contribution of the spherical correction is progressively overwhelmed by the planar Grahame term.

In terms of the stretched variables, the constant-charge boundary condition~\eqref{eq:CC} becomes:
\begin{equation}\label{eq:CC_stretched}
\left( \frac{\partial\phi}{\partial\xi} + \epsilon^{-1} \frac{\partial\phi}{\partial\rho} \right)_{\xi=0} = -\frac{2\lambda_D}{\ell_{GC}}.
\end{equation}
Substituting the regular perturbation series into Eq.~\eqref{eq:CC_stretched} and evaluating at the contact plane ($\xi=0$) yields the compiled boundary condition up to the first order in $\epsilon^{-1}$:
\begin{equation}\label{eq:CC_compiled}
\begin{split}
  \left.\frac{\partial\phi_0}{\partial\xi}\right|_{\xi=0}
  &+ \epsilon^{-1} \left( \left.\frac{\partial\phi_1}{\partial\xi}\right|_{\xi=0}
  + \left.\frac{\partial\phi_0}{\partial\rho}\right|_{\xi=0} \right) \\
  &+ \mathcal{O}(\epsilon^{-2}) = -\frac{2\lambda_D}{\ell_{GC}}.
\end{split}
\end{equation}

Substituting then the leading-order planar solutions~\eqref{eq:LC_planar} and the first-order spherical correction~\eqref{eq:phi1_prime_final} into Eq.~\eqref{eq:CC_compiled} eliminates the cross-derivative terms and yields:
\begin{equation}
-2\sinh\frac{\phi_s}{2} - 4\epsilon^{-1}\tanh\frac{\phi_s}{4} + \mathcal{O}(\epsilon^{-2}) = -\frac{2\lambda_D}{\ell_{GC}}.
\end{equation}
Multiplying both sides by $-1/2$ and substituting the definition of the parameter $\epsilon^{-1} = \lambda_D/R$, we establish that the charge-potential boundary relation up to the first order in curvature is given by:
\begin{equation}\label{LC}
\boxed{\frac{\lambda_D}{\ell_{GC}} \simeq \sinh\frac{\phi_s}{2} + \frac{2\lambda_D}{R}\tanh\frac{\phi_s}{4}}. \tag{LC}
\end{equation}
The first term on the right-hand side of Eq.~\eqref{LC} represents the classical Grahame relation for a planar electrostatic diffuse layer, while the second term captures the leading-order geometric contribution of the surface curvature.

We emphasize that Eq.~\eqref{LC} exactly reproduces the known asymptotic limits:
\begin{itemize}
  \item In the planar limit $R\to\infty$, the second term on the r.h.s. vanishes and we recover the exact Grahame relation \eqref{eq:grahame}.
  \item If $\phi_s$ is sufficiently small, $\sinh(\phi_s/2) \simeq \phi_s/2$ and $\tanh(\phi_s/4)\simeq \phi_s/4$, so Eq.~\eqref{LC} reduces to the linearized Eq.~\eqref{eq:contact_DH}.
  \item For large $\phi_s$, $\tanh(\phi_s/4)\to 1$, so the curvature correction saturates at $2\lambda_D/R$. Thus, Eq.~\eqref{LC} provides the high-potential asymptotic behavior where the curvature contribution becomes independent of the surface charge density.
\end{itemize}
Thus, one can expect that Eq.~\eqref{LC} will have validity beyond its particular restriction of small curvature.

We recall that the analytical form of this low-curvature relationship, Eq.~\eqref{LC}, was first introduced by \citet{ohshima.h:1982}. However, in their work, it was proposed merely as a mathematical ansatz to approximate early numerical solutions of the spherical Poisson--Boltzmann equation, constructed as an interpolation of the exact non-linear planar solution into a spherical geometry. To the best of our knowledge, a direct, mathematically rigorous derivation of Eq.~\eqref{LC} from first principles via a systematic regular expansion has been missing in the literature.

Therefore, our derivation bridges this theoretical gap. By employing the stretched boundary-layer coordinate $\xi$, we provide the rigorous proof that this ansatz is not merely a successful guess, but the mathematically exact, first-order regular asymptotic expansion of the spherical Poisson--Boltzmann equation for low curvatures ($\lambda_D/R \ll 1$).

\subsection{High Curvature Expansion (Small Sphere)}\label{subsec:high_curvature}

When the sphere is much smaller than the Debye length ($R \ll \lambda_D$), the standard thin-layer expansion fails because geometric divergence dominates over electrolyte screening near the surface. To resolve the boundary relation in this high-curvature regime, we keep the same dimensionless parameter $\epsilon = R/\lambda_D$, which now serves as our small parameter ($\epsilon \ll 1$), and employ the method of matched asymptotic expansions~\cite{vandyke1964}. We decompose the domain into an inner Laplacian region ($\rho = r/R = \mathcal{O}(1)$) and an outer Debye--H\"{u}ckel region ($\eta = \epsilon \rho = r/\lambda_D = \mathcal{O}(1)$).

In terms of the inner radial coordinate $\rho$, the dimensionless non-linear Poisson--Boltzmann equation for a symmetric electrolyte takes the form:
\begin{equation}\label{eq:PB_inner_coords}
\frac{1}{\rho^2}\frac{d}{d\rho}\left(\rho^2 \frac{d\phi^{(i)}}{d\rho}\right) = \epsilon^2 \sinh\phi^{(i)}.
\end{equation}
The electrostatic potential in both regions is formally expanded in powers of $\epsilon$:
\begin{align}
  \phi^{(i)}(\rho) &= \phi^{(i)}_0(\rho) + \epsilon\phi^{(i)}_1(\rho) + \mathcal{O}(\epsilon^2), \label{eq:series_phi_inner} \\
  \phi^{(o)}(\eta) &= \phi^{(o)}_0(\eta) + \epsilon\phi^{(o)}_1(\eta) + \mathcal{O}(\epsilon^2). \label{eq:series_phi_outer}
\end{align}
Substituting the inner expansion~\eqref{eq:series_phi_inner} into Eq.~\eqref{eq:PB_inner_coords} and collecting terms up to $\mathcal{O}(\epsilon)$ demonstrates that the screening term on the right-hand side is suppressed to $\mathcal{O}(\epsilon^2)$. Consequently, the governing equations for the first two orders in the inner zone reduce to uncoupled radial Laplace equations:
\begin{equation}\label{eq:Laplace_orders}
\frac{d}{d\rho}\left(\rho^2 \frac{d\phi^{(i)}_0}{d\rho}\right) = 0, \quad \text{and} \quad \frac{d}{d\rho}\left(\rho^2 \frac{d\phi^{(i)}_1}{d\rho}\right) = 0.
\end{equation}
Conversely, in the far-field outer region where the potential is small, the leading-order governing equation transforms into the linear spherical Debye--H\"{u}ckel equation:
\begin{equation}\label{eq:DH_outer}
\frac{1}{\eta^2}\frac{d}{d\eta}\left(\eta^2 \frac{d\phi^{(o)}_0}{d\eta}\right) = \phi^{(o)}_0.
\end{equation}

Dropping the screening term on the right-hand side of Eq.~\eqref{eq:PB_inner_coords} to maintain the Laplacian regime~\eqref{eq:Laplace_orders} is mathematically justified only as long as:
\begin{equation}\label{eq:linearity_criterion}
\epsilon^2 \sinh\phi_s \ll 1.
\end{equation}
Beyond this threshold, the Laplace approximation breaks down.

To evaluate the complete boundary relation up to $\mathcal{O}(\epsilon^2)$ without explicitly computing the separate spatial profiles for each order, we perform a first integration of the inner equations~\eqref{eq:Laplace_orders}, which shows that the constant $A = -\rho^2 d\phi^{(i)}/d\rho = -\left.d\phi^{(i)}/d\rho\right|_{\rho=1}$ is conserved across the inner domain up to $\mathcal{O}(\epsilon^2)$. Integrating from the surface ($\rho = 1$) into the intermediate overlap domain ($\rho \to \infty$, $\eta \to 0$) immediately yields the inner asymptotic behavior $\phi^{(i)}(\rho) \sim (\phi_s - A) + A/\rho + \mathcal{O}(\epsilon^2)$. In this overlap region, the far-field behavior of the inner Laplacian solution must match the small-scale limit of the outer screened Debye--H\"{u}ckel field, which is governed by the Yukawa-type solution of Eq.~\eqref{eq:DH_outer}: $\phi^{(o)}_0(\eta) = B \eta^{-1} e^{-\eta}$. Expanding this outer field for small distances yields $\phi^{(o)}_0(\eta) = B \eta^{-1} (1 - \eta + \mathcal{O}(\eta^2))$. Transformed back into the inner variable $\rho = \eta/\epsilon$, this expansion takes the form $\phi^{(o)}_0(\epsilon\rho) = B/(\epsilon\rho) - B + \mathcal{O}(\epsilon\rho)$. Matching the $1/\rho$ and constant terms between the inner and outer expansions gives $B = \epsilon A$ and $\phi_s - A = -\epsilon A$, which directly yields $\phi_s = A(1 - \epsilon) + \mathcal{O}(\epsilon^2)$.

This asymptotic matching condition transmits the distant screening physics back to the particle surface. Substituting $A = -\left.d\phi^{(i)}/d\rho\right|_{\rho=1}$ into $\phi_s = A(1 - \epsilon) + \mathcal{O}(\epsilon^2)$ and expanding to first order in $\epsilon$ yields:
\begin{equation}\label{eq:HC_compiled_derivative}
\left.\frac{d\phi^{(i)}}{d\rho}\right|_{\rho=1} = -\phi_s (1 + \epsilon) + \mathcal{O}(\epsilon^2).
\end{equation}
Applying the Neumann boundary condition $\left.\frac{d\phi^{(i)}}{d\rho}\right|_{\rho=1} = -\frac{2R}{\ell_{GC}}$ self-consistently closes the system and yields the surface potential $\phi_s$:
\begin{equation}\label{HC}
\boxed{\phi_s \simeq \frac{2R}{\ell_{GC}(1 + R/\lambda_D)}}. \tag{HC}
\end{equation}
The analytical expression~\eqref{HC} represents the contact theorem, being specific to the restriction of high curvature, and we do not expect it to be valid far beyond this range.

It is essential to emphasize the physical origin and significance of Eq.~\eqref{HC}, which is identical to Eq.~\eqref{eq:contact_DH}, as is easy to check. In the classical theory of strong electrolyte solutions~\cite{debye.p:1923}, a boundary relation of this form is derived by assuming a weak surface potential ($\phi_s \le 1$) to justify the linearization of the Poisson--Boltzmann equation. Crucially, no such weak-potential assumption is made in our derivation. Instead, Eq.~\eqref{HC} emerges as a strict asymptotic consequence of high surface curvature, which suppresses non-linearities. As a result, Eq.~\eqref{HC} remains exact even when the surface potential $\phi_s$ is quite large.

The above analysis provides a rigorous justification for the recent numerical findings of \citet{vinogradova.oi:2023,vinogradova.oi:2025}, whose calculations demonstrated that exact numerical solutions of the non-linear Poisson--Boltzmann equation unexpectedly recover the linear Debye--H\"uckel predictions even at high potentials. While our asymptotic analysis establishes the mathematical basis for this phenomenon, it also highlights its physical boundaries. Although no explicit constraints on the charge are imposed during the derivation of Eq.~\eqref{HC}, its applicability is strictly limited by the condition in Eq.~\eqref{eq:linearity_criterion}. Beyond this threshold, local non-linear screening dominates, forcing the potential profile back toward the non-linear Grahame limit (see the Appendix~\ref{app:equivalence} for more details). We postpone detailed discussion of the surface potential for this regime to a future paper.

\section{Numerical Results and Discussion}\label{sec:results}

It is of considerable interest to compare exact numerical data with our analytical theory and to determine the regimes of validity of asymptotic results.
Here we solve the non-linear spherical Poisson-Boltzmann equation~\eqref{eq:SPB} numerically with prescribed boundary conditions, using the collocation method~\cite{bader.g:1987}. The exact numerical solutions will be presented together with the results of the asymptotic approximations derived in Sec.~\ref{sec:expansions}.

\begin{figure}[h]
	\includegraphics[width=1.0\columnwidth]{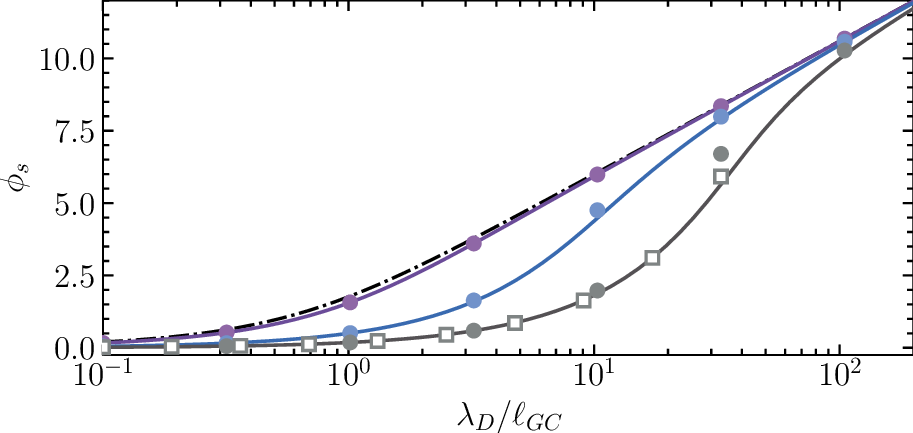}
	\caption{Surface potential $\phi_s$ as a function of $\lambda_D/\ell_{GC}$ computed for $\epsilon^{-1}=\lambda_D/R = 0.2, 3,$ and $10$ (solid curves from top to bottom). The dash-dotted curve is calculated from the flat Grahame relation~\eqref{eq:grahame}. Open and filled symbols show the predictions of Eqs.~\eqref{HC} and~\eqref{LC}, respectively. }
	\label{fig:ct_num_log}
\end{figure}

Figure~\ref{fig:ct_num_log} shows the numerical results (solid curves) for the surface potential $\phi_s$ as a function of the dimensionless charge $\lambda_D/\ell_{GC}$ on a lin-log scale. The calculations are made for several fixed $\epsilon^{-1}=\lambda_D/R$ in the range from $0.2$ to $10$. Also included is the (dash-dotted) curve calculated from exact Eq.~\eqref{eq:grahame} for a flat wall. It can be seen that when $\lambda_D/\ell_{GC} \to 0$ all the curves vanish. On increasing $\lambda_D/\ell_{GC}$ they separate and initially exhibit a linear growth (which appears concave in this lin-log scale). It is also seen that their slope reduces as $\epsilon^{-1}$ increases. Crucially, when $\epsilon^{-1} = 0.2$, the numerical curve is so close to the limiting case of a flat interface that, at first glance, the use of a low-curvature expansion appears unnecessary. For larger values of $\epsilon^{-1}$, deviations from the flat Grahame relation~\eqref{eq:grahame} become more pronounced, shifting the surface potential toward lower values than predicted for a planar wall. On increasing $\lambda_D/\ell_{GC}$ further $\phi_s$ augments more rapidly and when $\lambda_D/\ell_{GC} \to \infty$ all curves virtually merge into a single straight line. In this limit, the flat Grahame relation~\eqref{eq:grahame} applies for any curvature and $\phi_s$ grows weakly logarithmically.

We next compare these numerical results with the predictions of the low-curvature relationship~\eqref{LC}. For $\epsilon^{-1} = 0.2$, the analytical predictions expectedly coincide with the numerical curve across the entire range of parameters. To see if~\eqref{LC} performs beyond its formal range of applicability, we tested the equation for the other two curves. Remarkably, it demonstrates a very wide range of validity even for larger curvatures. Even at a moderate $\epsilon^{-1} = 3$, Eq.~\eqref{LC} closely follows the numerical curve, yielding only a very slight, practically undiscernible, overestimation around $\lambda_D/\ell_{GC} \simeq 10$. For a highly curved system ($\epsilon^{-1} = 10$), the regular expansion~\eqref{LC} remains highly accurate except in the range of $\lambda_D/\ell_{GC}$ from ca. $10$ to $10^2$, where the overestimation of the surface potential by Eq.~\eqref{LC} becomes clearly visible, but still remains rather small.

To understand how and why relationship~\eqref{LC} captures the numerical trends across all curvatures, we analyze its asymptotic behavior. At small $\lambda_D/\ell_{GC}$, Eq.~\eqref{LC} naturally reduces to the spherical Debye-Hückel solution (see Sec.~\ref{subsec:low_curvature}), which clearly explains why the initial slope of the curves decreases as $\lambda_D/R$ increases. In the opposite limit of very high $\lambda_D/\ell_{GC}$ at  fixed $\epsilon^{-1}$, Eq.~\eqref{LC} reduces to:
\begin{equation}
\phi_s \simeq 2 \ln \left( \frac{2\lambda_D}{\ell_{GC}} - \frac{4\lambda_D}{R} \right).
\end{equation}
Thus, the curvature introduces a radius-dependent correction directly within the Grahame logarithm. On a lin-log scale, this shifts the onset of the asymptotic regime and explains why the curves approach the planar Grahame limit~\eqref{eq:grahame} with different slopes before finally merging.
Consequently, the equation originally derived for low curvature remains remarkably accurate across all regimes, as it successfully interpolates between the two exact limits: it naturally reduces to the spherical Debye-Hückel solution in the wide linear regime of high curvature and converges to the Grahame equation at high charges. While it becomes effectively redundant for low-curvature systems where the classic Grahame solution suffices, it serves as a highly successful approximation for intermediate curvatures.

We now turn to the linear relationship~\eqref{HC} derived in Sec.~\ref{subsec:high_curvature} for high curvatures. The calculations from Eq.~\eqref{HC} are also included in Fig.~\ref{fig:ct_num_log} and compared with the lowermost numerical curve ($\lambda_D/R = 10$). The fit is excellent for $\lambda_D/\ell_{GC} \lesssim 40$ (or $\phi_s \lesssim 6$), but at higher values, the surface potential predicted by Eq.~\eqref{HC} increases more rapidly than the true numerical one.
The point is that at this limit of huge charge, local non-linear screening eventually dominates the electrostatic response. Even though no explicit constraints on the charge were imposed during the derivation of Eq.~\eqref{HC}, the Gouy-Chapman length $\ell_{GC}$ becomes much smaller than the sphere radius $R$. As a result, the non-linear diffuse layer becomes extremely thin compared to the surface curvature, and the true surface potential exhibits a transition into the logarithmic scaling, causing Eq.~\eqref{HC} to inevitably overestimate $\phi_s$. However, we conclude that due to the strong geometric regulation effect, the range of applicability for Eq.~\eqref{HC} is remarkably wide, accurately capturing the boundary electrostatics up to quite large dimensionless charges.

\begin{figure}[h]
	\includegraphics[width=1.0\columnwidth]{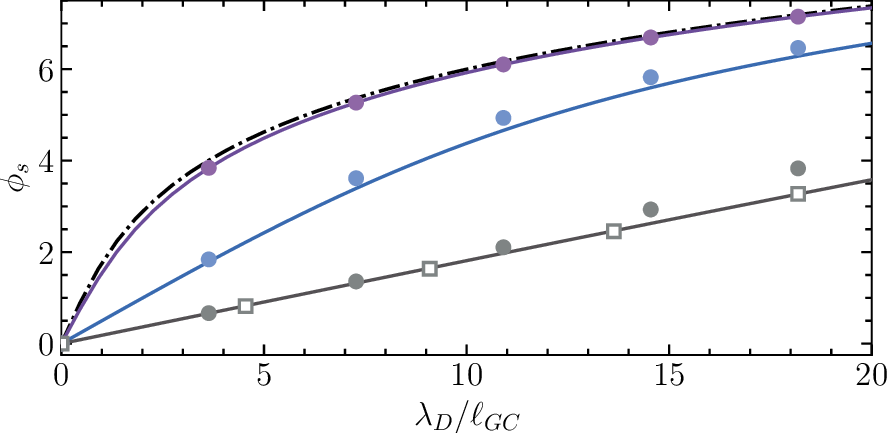}
	\caption{The same data as in Fig.~\ref{fig:ct_num_log}, replotted to highlight the low-charge regime up to $\lambda_D/\ell_{GC} = 20$ using a linear scale for both axes.}
	\label{fig:ct_num}
\end{figure}

To examine the low- and moderate-charge regions more closely, the initial portions of the curves up to $\lambda_D/\ell_{GC} = 20$ are presented in Fig.~\ref{fig:ct_num}. In this regime, the regular expansion~\eqref{LC} provides a reasonable approximation for all three curves, exhibiting only a slight overestimation for the larger curvatures ($\epsilon^{-1} = 3$ and $10$). Crucially, Eq.~\eqref{HC} completely coincides with the numerical data for the highly curved sphere ($\epsilon^{-1} = 10$). This perfect agreement validates our theoretical finding that the geometric scaling of the spherical surface automatically regulates the boundary electrostatics, forcing the exact solution to obey the linearized framework at low-to-moderate surface charges.

To place these findings in a practical context, it is instructive to map the dimensionless parameters onto characteristic experimental scales. It is now convenient to use the molar concentration $c_\infty$ [mol/l] related to the number density [m$^{-3}$] as $n_\infty = N_A \times 10^3 \times c_\infty$, where $N_A$ is Avogadro's number. Such highly curved geometries with $\epsilon^{-1} = 10$, corresponding to the lowermost curve in Fig.~\ref{fig:ct_num}, are directly relevant to ultrasmall functionalized nanoparticles, molecular-sized quantum dots, and stable spherical micelles. For a typical $1$:$1$ aqueous electrolyte solution at room temperature with a bulk concentration of $c_{\infty} = 10^{-3}$~$\text{mol/l}$ ($\lambda_D \simeq 9.6$~nm), a highly curved nanoparticle of radius $R = 0.96$~nm bearing a surface charge density of $\sigma \simeq 0.076$~$\text{C/m}^2$ corresponds to $\lambda_D/\ell_{GC} = 20$. Under these specific conditions, the classic flat-wall Grahame relation~\eqref{eq:grahame} yields $\phi_s \simeq 7.4$ ($\Phi_s \simeq 190$~mV), completely failing to account for the strong curvature, while the standard flat-wall Debye–Hückel approximation breaks down entirely, predicting an unphysical value of $\phi_s = 40$ ($\Phi_s \simeq 1030$~mV). In contrast, the classic spherical Debye–Hückel approximation yields $\phi_s \simeq 3.64$ ($\Phi_s \simeq 93.4$~mV). Our rigorous high-curvature relationship~\eqref{HC} reduces exactly to this classic linear solution, perfectly matching the exact numerical solution of the full non-linear Poisson--Boltzmann equation. This directly demonstrates that for such highly curved spheres, the electrostatic response is so strongly regulated by geometry that it strictly follows the linearized framework even at these exceedingly high surface charge densities.

\begin{figure}[h]
	\includegraphics[width=1.0\columnwidth]{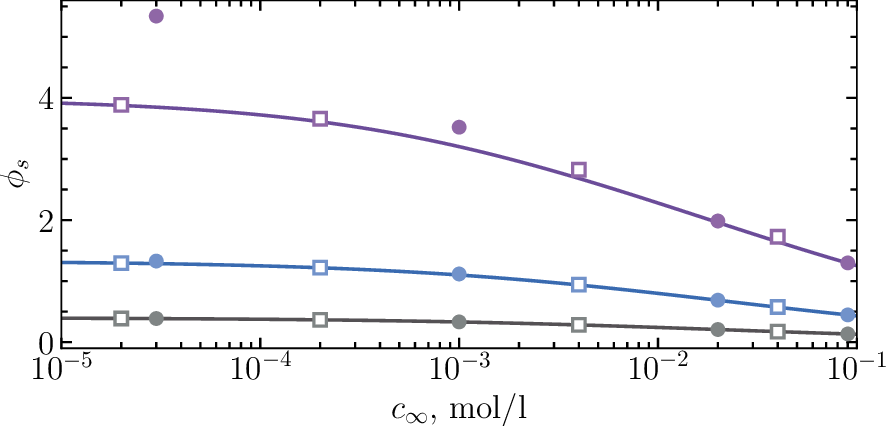}
	\caption{Surface potential $\phi_s$ as a function of the bulk salt concentration $c_{\infty}$ computed for a spherical particle of radius $R = 2$\,nm using $\ell_{GC} = 1$, $3$, and $10$\,nm (solid curves from top to bottom). Open and filled symbols show predictions of Eqs.~\eqref{HC} and \eqref{LC}, respectively.}
	\label{fig:ct_salt}
\end{figure}

To illustrate these trends in variables directly accessible to experiments, in Fig.~\ref{fig:ct_salt} we plot the numerical surface potential $\phi_s$ of a particle as a function of the bulk salt concentration $c_{\infty}$ (solid curves). For this example, we set $R = 2$\,nm and show three curves corresponding to fixed charge densities with $\ell_{GC} = 1$ (high charge), $3$ (moderate charge), and $10$\,nm (low charge). It can be seen that the surface potential decreases monotonically as $c_{\infty}$ increases, which is consistent with enhanced electrostatic screening. This attenuation is more pronounced for the uppermost curve ($\ell_{GC} = 1$\,nm), although surface potentials retain finite values over the entire concentration range. The intermediate curve ($\ell_{GC} = 3$\,nm) exhibits a variation from moderate down to small potentials, whereas the lowermost one ($\ell_{GC} = 10$\,nm) remains entirely within the low-potential regime.

Also included in Fig.~\ref{fig:ct_salt} are the surface potentials calculated from Eqs.~\eqref{HC} and \eqref{LC}, shown by open and filled symbols, respectively. The purpose is to see how these two expansions divide their spheres of influence as salt concentration changes. At $c_{\infty} \leq 2.3 \times 10^{-4}$\,mol/l (large $\lambda_D$, such that $\epsilon = R/\lambda_D \ll 1$), the system enters the high-curvature regime. At larger concentrations it begins to transition smoothly toward, although it does not reach with these parameters, the low-curvature regime. However, it can be seen that both Eq.~\eqref{LC} and Eq.~\eqref{HC} fit the two lower numerical curves (low and moderate charge) perfectly, i.e., have validity beyond these particular models. Indeed, at low and moderate potentials ($\phi_s \lesssim 1$, corresponding to $\ell_{GC} = 3$ and $10$\,nm), Eq.~\eqref{LC} algebraically reduces to the linearized spherical Debye--H\"{u}ckel relation (identical to Eq.~\eqref{HC}), which is exact for arbitrary $\epsilon$. This explains why both expansions capture the geometric factor $(1 + \epsilon)^{-1}$ across the entire concentration range.
 The uppermost curve (high charge) is different. Here, Eq.~\eqref{HC} captures the entire concentration range: at low salt concentrations, non-linearity is suppressed by high curvature ($\epsilon \ll 1$, satisfying $\epsilon^2 \sinh\phi_s \ll 1$), whereas at high salt concentrations, enhanced screening reduces $\phi_s$ into the weakly non-linear regime where Eq.~\eqref{HC} remains accurate. By contrast, Eq.~\eqref{LC} significantly overestimates the numerical data at $c_{\infty} \leq 10^{-3}$\,mol/l. While Eq.~\eqref{LC} seamlessly reduces to the exact Debye--H\"{u}ckel limit at low potentials regardless of $\epsilon$ (explaining its success for the two lower curves), at high surface potentials its structure relies on perturbing the planar Grahame relation to $\mathcal{O}(\epsilon^{-1})$. Extending this quasi-planar expansion into the high-curvature regime ($\epsilon \ll 1$), where spherical geometry actually suppresses planar non-linearity, causes the neglected $\mathcal{O}(\epsilon^{-2})$ curvature terms to grow rapidly and severely overestimate $\phi_s$.

\begin{figure}[h]
	\includegraphics[width=1.0\columnwidth]{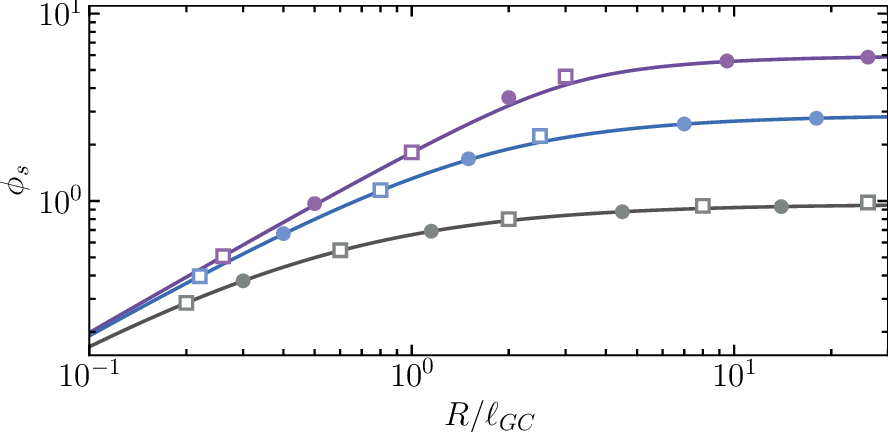}
	\caption{Surface potential $\phi_s$ vs particle radius $R/\ell_{GC}$ computed at $c_{\infty} = 10^{-3}$\,mol/l using $\lambda_D/\ell_{GC}=0.5$, $2$, and $10$ (solid curves from bottom to top). Open and filled symbols show predictions of Eqs.~\eqref{HC} and \eqref{LC}, respectively.
}
	\label{fig:ct_R}
\end{figure}

These results refer to fixed $R$ and $\ell_{GC}$ and varying $c_{\infty}$. If we fix $\lambda_D/\ell_{GC}$ and $c_{\infty}$, and vary $R/\ell_{GC}$, we move to the situation described by Fig.~\ref{fig:ct_R}. The calculations are performed for $c_{\infty} = 10^{-3}$\,mol/l, which yields $\lambda_D \simeq 9.6$\,nm, using $\lambda_D/\ell_{GC} = 0.5$ (weak charge, $\ell_{GC} \simeq 20$\,nm), $2$ (moderate charge, $\ell_{GC} \simeq 5$\,nm), and $10$ (strong charge, $\ell_{GC} \simeq 1$\,nm). On this log--log scale, all computed curves monotonically increase with the radius, capturing the geometric transition from highly curved spherical particles to the flat wall regime. As discussed in Sec.~\ref{sec:framework}, when $R/\ell_{GC} \to 0$ all the curves vanish. At very small $R/\ell_{GC}$, they follow a common linear scaling ($\phi_s \simeq 2R/\ell_{GC}$), and upon increasing $R/\ell_{GC}$ they separate as curves with smaller $\lambda_D/\ell_{GC}$ experience screening earlier and bend downward sooner. On increasing $R/\ell_{GC}$ further, $\phi_s$ increases more slowly, and when $R/\ell_{GC} \gg 1$ all curves saturate to a respective plateau described by the Grahame equation~\eqref{eq:grahame}. To provide concrete values, the saturation surface potential is $\phi_s \simeq 6.0$ (strong charge), $2.89$ (moderate charge), and $0.96$ (weak charge).

Crucially, Fig.~\ref{fig:ct_R} highlights the distinct physical validity domains of the analytical frameworks by comparing the numerical data with the predictions of Eqs.~\eqref{HC} and~\eqref{LC}. For the weakly charged particle ($\lambda_D/\ell_{GC} = 0.5$, lowermost curve), Eq.~\eqref{HC} accurately reproduces the numerical curve across the entire range of radii, since $\phi_s$ is below unity everywhere and the system remains within the linear Debye--H\"{u}ckel regime for arbitrary $\epsilon$. As the surface charge increases ($\lambda_D/\ell_{GC} = 2$ and $10$), the validity of Eq.~\eqref{HC} becomes confined to the small-radius, high-curvature domain where $\epsilon$ is sufficiently small to satisfy criterion~\eqref{eq:linearity_criterion}. A crossover between two regimes occurs for some values $R/\ell_{GC} = \mathcal{O}(2-4)\sim \ln (\epsilon^{-1})$ (see the Appendix). Specifically, for the strongly charged sphere ($\lambda_D/\ell_{GC} = 10$, uppermost curve), Eq.~\eqref{HC} holds remarkably well up to $R/\ell_{GC} \simeq 3.0$ (which yields $\phi_s \simeq 4.6$ and $\epsilon \simeq 0.3$, or $\epsilon^{-1} \simeq 3.3$), before starting to overestimate the exact potential as non-linear screening sets in. Thus, high curvature, even when coupled to strong surface charge, keeps the linear~\eqref{HC} relation accurate well into the finite-potential regime. By contrast, the non-linear Eq.~\eqref{LC} fits the exact numerical curves surprisingly well across the entire range of radii shown. This result is startling since the local dimensionless curvature varies continuously by three orders of magnitude along each profile. For instance, on the uppermost curve, $\epsilon^{-1}$ drops from $10^2$ at the left boundary down to $0.33$ at the right edge. The good agreement observed highlights a successful crossover between the two asymptotes rather than universal validity under arbitrary conditions, and is also a direct consequence of the specific parameter space used in Fig.~\ref{fig:ct_R}. Indeed, the limitations of Eq.~\eqref{LC} are clearly manifested in the high-curvature regime in Figs.~\ref{fig:ct_num_log}--\ref{fig:ct_salt}, where our singular expansion~\eqref{HC} achieves an excellent alignment with the exact Poisson--Boltzmann solution.

\section{Final Remarks}\label{sec:concl}

In summary, the dual-asymptotic approach developed in this work provides a rigorous analytical description of the boundary electrostatics for charged spherical particles across both curvature extremes. By combining a regular asymptotic expansion for weakly curved surfaces with a matched asymptotic treatment for highly curved geometries, we obtain explicit closed-form charge--potential relations, thereby reducing the reliance on numerical computations in these limits. The resulting expressions are simple and accurate. By bypassing the need for case-specific numerical routines, they offer a direct, practical analytical foundation for various applications across colloid and interface science.

We are not the first to obtain the low-curvature contact relationship. \citet{ohshima.h:1982} proposed it as a mathematical ansatz interpolating between the exact non-linear planar solution and the spherical geometry. Our work clarifies that this expression constitutes the exact first-order asymptotic result of the spherical Poisson--Boltzmann equation and defines the range of its validity. Moreover, our regular perturbation expansion offers a direct route to the full potential profile and the effective charge, which would not be possible using the approach by \citet{ohshima.h:1982}. The regular expansion method is also considerably simpler than the singular perturbation approach of \citet{natarajan.r:1984}, where matching inner and outer regions prevented expressing the first-order curvature corrections in terms of elementary functions. Our work is in progress on this.

The compact high-curvature expression derived in our work, which has not appeared in closed form in the literature before, justifies and provides considerable insight into the geometric deactivation of mean-field non-linearity, validating the classical Debye--H\"{u}ckel result at high potentials. Certain aspects of our results for highly curved particles warrant the further comments given below.

Beyond its practical utility, our matched asymptotic solution for the high-curvature regime yields a clear physical insight, offering a rigorous mathematical justification for a classic concept rooted in the foundational work of~\citet{debye.p:1923}. When dealing with the point-charge singularity, classical electrolyte theory introduces the finite particle radius $R$ as an explicit physical cutoff---the distance of closest approach---to prevent the Coulomb field from exploding. While a point-like limit within the full non-linear Poisson--Boltzmann theory is traditionally expected to trigger a divergence of the Boltzmann factor---a short-range pathology first diagnosed by~\citet{onsager.l:1933}---we demonstrate that the mathematical structure of the equations inherently tames this short-range divergence. At fixed surface charge density, the three-dimensional spatial dilution of the field forces the dimensionless boundary potential to scale down to zero ($\phi_s \to 0$) as the particle shrinks, while for finite dimensionless potentials ($\phi_s = \mathcal{O}(1)$) the geometric prefactor $\epsilon^2 = (R/\lambda_D)^2$ suppresses the non-linear screening term. This formal convergence proves that the linearized Debye--H\"{u}ckel approximation is the mathematically exact first-order asymptotic limit of the full non-linear problem under high curvature. Consequently, our work provides a validation of the classical Debye--H\"{u}ckel approach for isolated spherical particles, demonstrating that introducing a geometric boundary radius is sufficient to restore linearity without triggering point-charge singularities at extreme curvatures.

The practical utility of our geometrically driven linearization becomes especially important when analyzing real-world soft-matter and biomedical systems operating in the regime of ultra-high curvature paired with high surface charge density~$\sigma$. Prominent examples include functionalized ultrasmall silica nanoparticles (with a radius of $1\text{--}2~\mathrm{nm}$) used in targeted cancer theranostics, quantum dots, dense spherical surfactant micelles, low-generation charged dendrimers, and even inorganic ions. Conventionally, the electrostatic profiles of these highly charged, nanoscale objects have been notoriously difficult to evaluate analytically, often forcing researchers to rely on computationally expensive molecular dynamics or heavy numerical solvers. Our results justify why these seemingly extreme systems remain well-behaved: as the physical scale approaches the nanometer or sub-nanometer limit, the three-dimensional geometry naturally suppresses the mean-field electrostatic collapse. This validates the use of our streamlined analytical relationship for precise predictions of their colloidal stability and transport behavior.

In a broader context, these findings provide a missing analytical foundation for the physics of complex soft architectures at finite salt concentrations. Historically, theoretical descriptions of highly curved systems---such as microgels, dendrimers, and spherical polyelectrolyte brushes---have frequently challenged standard models, forcing a difficult choice between heavy numerics and \textit{ad hoc} linear approximations. Our matched asymptotic derivation formally resolves this tension from first principles. By proving that the boundary electrostatics naturally sheds its non-linear character as the curvature radius vanishes, this approach establishes a rigorous theoretical foundation that smoothly bridges the gap between fundamental electrostatics and broader applications in biophysics and colloid science.

\appendix*
\section{Geometric Threshold for Non-Linear Screening}\label{app:equivalence}

To link the threshold in Eq.~\eqref{eq:linearity_criterion} with the geometric ratio of the particle radius to the Gouy--Chapman length, we evaluate the screening term at the particle surface ($\rho = 1$) in the high-curvature, strongly charged regime. At the surface of a small sphere ($\epsilon = R/\lambda_D \ll 1$), the leading-order potential reduces to $\phi_s \simeq 2R/\ell_{GC}$ (Eq.~\eqref{eq:phil_sp}). For large potentials ($\phi_s \gg 1$), where $\sinh\phi_s \simeq e^{\phi_s}/2$, the condition $\epsilon^2 \sinh\phi_s \ll 1$ takes the form
\begin{equation}\label{app:eq:screening_scale}
\frac{1}{2}\epsilon^2 \exp\!\left(\frac{2R}{\ell_{GC}}\right) \ll 1.
\end{equation}
Taking the logarithm of the crossover condition $\epsilon^2 \exp(2R/\ell_{GC}) \sim 1$ immediately yields the scaling for the breakdown threshold:
\begin{equation}\label{app:eq:log_criterion}
\frac{R}{\ell_{GC}} \sim \ln\!\left(\epsilon^{-1}\right) = \ln\!\left(\frac{\lambda_D}{R}\right).
\end{equation}
Because $\epsilon$ is small in the high-curvature regime, the logarithmic factor $\ln(\lambda_D/R)$ is strictly larger than unity (typically ranging from $\sim 2.3$ at $\epsilon^{-1} = 10$ up to $6$ or more for ultra-small ions in dilute electrolytes). Thus, Eq.~\eqref{app:eq:log_criterion} demonstrates that the breakdown of the inner Laplacian regime occurs not merely when $R \sim \ell_{GC}$, but only once the particle radius $R$ substantially exceeds the Gouy--Chapman length $\ell_{GC}$ by this logarithmic factor. This logarithmic buffer explains why Eq.~\eqref{HC} remains accurate well into the strongly charged regime ($R/\ell_{GC} > 1$). Beyond this threshold, exponential screening overwhelms the small geometric prefactor $\epsilon^2$, causing the exact potential to deviate downward from the high-curvature relation~\eqref{HC} toward the non-linear Grahame limit.

This scaling behavior is directly confirmed by our numerical results. In the variable-charge representation of Fig.~\ref{fig:ct_num_log} (where $R$ is fixed), the lowermost curve ($\epsilon^{-1} = 10$, i.e., $\epsilon = 0.1$) yields $\ln(\epsilon^{-1}) \simeq 2.3$. Indeed, Eq.~\eqref{HC} tracks the exact numerical solution closely up to $\lambda_D/\ell_{GC} \simeq 35\text{--}40$, beyond which exponential screening triggers a pronounced deviation from the inner Laplacian approximation. Similarly, in the variable-radius framework of Fig.~\ref{fig:ct_R} (where $\ell_{GC}$ and $\lambda_D$ are fixed, with $c_{\infty} = 10^{-3}$\,mol/l), the same criterion governs the departure from Eq.~\eqref{HC} at strong surface charges ($\lambda_D/\ell_{GC} = 10$, uppermost curve), where deviations appear around $R/\ell_{GC} \simeq 2.5\text{--}3$ (where $\epsilon \simeq 0.25$--$0.3$ and $\ln(\epsilon^{-1}) \simeq 1.2$--$1.4$). By contrast, for moderate and weak charges ($\lambda_D/\ell_{GC} = 2$ and $0.5$), the factor $(1 + \epsilon)^{-1}$ in Eq.~\eqref{HC} keeps $\phi_s$ moderate across all radii, preventing exponential blow-up and extending the accuracy of Eq.~\eqref{HC} over the entire range shown.

\begin{acknowledgments}
We have benefited from discussions with E.~S.~Asmolov. This work was supported by the Ministry of Science and Higher Education of the Russian Federation.
\end{acknowledgments}

\section*{DATA AVAILABILITY}

The data that support the findings of this study are available within the
article.

\section*{AUTHOR DECLARATIONS}

The authors have no conflicts to disclose.

\end{document}